\documentclass[9pt]{article}
\usepackage[utf8x]{inputenc}
\usepackage[T1]{fontenc}
\usepackage{amsmath}
\usepackage{amsthm}
\usepackage{amsfonts}
\usepackage{amssymb}
\usepackage{amscd}
\usepackage{bm}
\usepackage{dsfont}
\usepackage{enumitem}
\usepackage{graphicx}
\usepackage{subfigure}
\usepackage{algorithm}
\usepackage{algorithmic}

\usepackage{xcolor}
\definecolor{darkgreen}{rgb}{0,0.7,0}
\definecolor{darkred}{rgb}{0.6,0,0}
\usepackage[pdftex,
            bookmarks=true,
            bookmarksnumbered=true,
            pdfpagemode=UseOutlines,
            pdfstartview=FitH,
            pdfpagelayout=OneColumn,
            colorlinks=true,
            linkcolor=darkred,
            citecolor=blue,
            pdfborder={0 0 0}]{hyperref}
\usepackage{comment}
\usepackage{tikz}
\usetikzlibrary{arrows,calc,matrix}
\tikzstyle{block}=[draw opacity=0.7,line width=1.4cm]

\theoremstyle{plain}

\theoremstyle{definition}

\numberwithin{theorem}{section}
\numberwithin{defn}{section}
\numberwithin{equation}{section}

\newcommand{\abs}[1]{\left\lvert#1\right\rvert}

\newcommand{\real}{\mathbb{R}}

\newcommand{\CO}{\mathcal{O}}

\newcommand{\myobot}{\mathop{\bigcirc\kern-11.5pt\perp}\nolimits}

\title{A Short Note on Zero-error Computation for Algebraic Numbers by IPSLQ
\thanks{Key words: zero-error computation, 
algebraic number, integer relation. This is a modified version of \cite{FengChenWu2013}.}}

\author{Yong Feng \ \ \ \ Jingwei Chen
\ \ \ \  Wenyuan Wu\\
Chongqing Key Laboratory of Automated Reasoning and Cognition,\\
Chongqing Institute of Green and Intelligent Technology, CAS}

\date{
}

\begin{document}

\maketitle

\section{Introduction}
Being exact, symbolic computation is usually inefficient due to the well-known problem of intermediate swell. 
Being efficient, numerical computation only gives approximate results. 
For both efficiency and reliability,
an idea that obtains exact results by using approximate
computing has been of interest. 
We call such methods \emph{zero-error computation}.

The output of \emph{zero-error algorithms} are exact expressions,
but the intermediate process (partially) uses
appropriate numerical methods, so that they are symbolic-numeric. At the end of
these algorithms the errors become zero (i.e., exact results) by certain gap 
theorem from background knowledge, 
though errors appear at (almost) every step in these algorithms. 
For instance, applying the algorithm
presented in \cite{ZhangFeng2007}, one can recovery the exact value
of a rational number from its approximation; more generally, the
algorithm in \cite{KannanLenstraLovasz1988} is to reconstruct
algebraic numbers.
In many zero-error computation algorithms, such as
reconstruction algebraic number \cite{KannanLenstraLovasz1988},
polynomial factorization \cite{vanHoeij2002, WuChenFeng2014}, etc.,
the problem to be solved is finally converted to finding an integer relation.

The PSLQ algorithm is one of the most popular algorithm for finding
nontrivial integer relations for several real numbers. 
Although it has been theoretically proved that the PSLQ algorithm \cite{FergusonBaileyArno1999}
is to some extent equivalent to the HJLS algorithm \cite{HastadJustLagariasSchnorr1989}, under
the exact real arithmetic computational model (see, e.g., \cite{Meichsner2001}), the PSLQ algorithm
seems more practical. The problem of finding the minimal polynomial from an approximation $\overline{\alpha}$
of a $d$ degree algebraic number $\alpha$, equivalent to finding an integer relation for  the vector $(1,\alpha,\ldots, \alpha^d)$,
was first solved in \cite{KannanLenstraLovasz1988} by using the celebrated LLL algorithm \cite{LenstraLenstraLovasz1982}.
This routine has been recently improved in \cite{vanHoeijNovocin2012}.
Naturally, the PSLQ algorithm is applicable to the algebraic number reconstruction problem as well \cite{QinFengChenZhang2012}.

Given an approximation to $\alpha$, a degree bound $d$ and an upper bound $M$ on its height, if we do not know the
exact degree of the algebraic number in advance, then no matter whether one uses PSLQ or LLL,
one has to search an integer relation for the vector $(1,\alpha,\ldots, \alpha^i)$ from $i=2, 3,\ldots$
until the degree bound $d$. Hence, if the complexity of an algorithm for finding an integer relation is $\CO(P(n, M))$
for an $n$-dimensional vector,
then the complexity of the minimal polynomial algorithm, based on the integer relation finding algorithm,
is $\CO(d\cdot P(d, M))$.
Our main contribution in the present work is to give the incremental PSLQ algorithm (IPSLQ), based on which,
the corresponding algebraic number reconstruction algorithm has the complexity only $\CO(P(d, M))$, even
though we do not know the exact degree of the algebraic number.

\section{The Incremental PSLQ Algorithm}

The main difference between IPSLQ (Algorithm \ref{algo:mpslq}) and PSLQ is the following:
PSLQ considers $x_1,\ldots,x_n$; IPSLQ considers
$x_{i}, \ldots, x_n$, if the vector $(x_{i}, \ldots, x_n)$ has no relation with $2$-norm less than $M$
(see Step \ref{step:mpslq:trmntn}) then add $x_{i-1}$ to the left.

When reconstructing the minimal polynomial, we apply IPSLQ with input as
 $(x_{1}, \ldots, x_n)=(\alpha^{n-1}, \ldots,\alpha, 1)$. Now,
 if the vector $(x_{i}, \ldots, x_n)$ has no relation with $2$-norm less than $M$,
the results of previous iterations can be reused in the next iteration. However,
the traditional methods can not use the information produced by the previous iterations.
Therefore, the complexity of IPSLQ for minimal polynomial without knowing the degree
is only $\CO(P(d, M))$, which is the same as PSLQ for minimal polynomial with knowing the degree.

\begin{algorithm}[H]
\caption{(IPSLQ).}\label{algo:mpslq}
\begin{algorithmic}
\REQUIRE A  vector $\bm x=(x_1,\cdots,x_n)\in\real^n$ with $x_i\neq 0$ and a positive number $M$.
\ENSURE Either return an integer relation for $\bm x$, or return ``$\lambda_1(\bm x)>M$''.
\end{algorithmic}
\begin{enumerate}[leftmargin=*, topsep=0pt,partopsep=0pt,itemsep=0pt,parsep=0pt]
\item\label{step:mpslq:ini} Compute $H_x\in\real^{n\times (n-1)}$.
Set $H:=H_x$, $A:= I_n$ and $B:= I_n$. Size-reduce $H$ and update $A$ and $B$.
\item\label{step:mpslq:grdl} For $k$ from n-1 to 1 do
\begin{enumerate}[leftmargin=*, topsep=0pt,partopsep=0pt,itemsep=0pt,parsep=0pt]
\item\label{step:mpslq:while} While $h_{n-1, n-1}\neq 0$ do
\begin{enumerate}[leftmargin=*, topsep=0pt,partopsep=0pt,itemsep=0pt,parsep=0pt]
\item Choose $r$ such that $\gamma^r\abs{h_{r,r}} = \max_{j\in\{k,\cdots,n-1\}}\left\{\gamma^j \abs{h_{j,j}}\right\}$.
\item Swap the $r$-th  and the $(r+1)$-th rows of $H$ and update $A$ and $B$.
\item If $r<n-1$ then  update $H$ to L-factor of $H$.
\item Size-reduce $H$ and update $A$ and $B$.
\item\label{step:mpslq:trmntn} If $\max_{j\in\{k,\cdots, n-1\}}\abs{h_{j,j}}<1/M$  then do the following:
 If $k>1$ then go to Step \ref{step:mpslq:grdl}; Else return ``$\lambda_1(\bm x)>M$''.
\end{enumerate}
\item Return the last column of $B$.
\end{enumerate}
\end{enumerate}
\end{algorithm}

\section{Experiments}

The following experiments are preliminary and  to compare the
performance between traditional PSLQ and IPSLQ for minimal polynomial reconstruction.

Consider approximations of
$\alpha = 3^{1/s}+2^{1/t}$ with 500 decimal digits. Running these experiments in Maple
15 with \texttt{Digits :=500} gives a preliminary experimental results in Table \ref{tab:MPSLQvsPSLQ}.
Note that here \texttt{Digits :=500} may not be necessary for many examples (see \cite{QinFengChenZhang2012}
for the a detailed error control).
In Table \ref{tab:MPSLQvsPSLQ},
the input degree bound and height bound in these tests are $d$ and $M+1$;
the exact degree and height of $\alpha$ are $d-1$ and $M$, respectively.
All these experimental results are obtained by using a Windows 7 (32 bits mode) PC with AMD Athlon  II X4 645
processor (3.10~GHz) and 4~GB memory.

\begin{table}[!ht]\centering
\begin{tabular}{cccccrrr}
  \hline
  No. & $s$ & $t$ & $d$ & $M$ &$T_{IPSLQ}$ & $T_{PSLQ}$ & $\frac{T_{PSLQ}}{T_{IPSLQ}}$\\\hline
  1 & 2 & 2 & 5& 10& 0.08  &  0.16&2.00\\
  2 & 2 & 3 & 7& 36&   0.16&   0.64&4.00\\
  3 & 3 & 3 & 10& 125&  0.89&   5.34&6.00\\
  4 & 3 & 4 & 13& 540&  3.14&   21.34&6.79\\
  5 & 2 & 7 & 15& 5103&  6.91&   45.91&6.64\\
  6 & 3 & 6 & 19& 10278& 23.37&   144.11&6.17\\
  7 & 4 & 5 & 21& 11160& 32.73&   249.54&7.62\\
  8 & 5 & 5 & 26& 57500& 78.95&   838.99&10.63\\
  9 & 5 & 6 & 31& 538380& 186.28&   2089.87&11.22\\
  10 & 6 & 6 & 37& 4281690& 421.94&   4313.99&10.22\\
  \hline
\end{tabular}\caption{The comparison of IPSLQ and PSLQ  for minimal polynomial}
\label{tab:MPSLQvsPSLQ}
\end{table}

Note that there exists a built-in function \texttt{IntegerRelations:-PSLQ} in Maple 15, but
for the comparison in Table \ref{tab:MPSLQvsPSLQ}, we implement the PSLQ algorithm by ourselves. The
reasons we do not use the built-in function
is that there does not exist a height parameter in the built-in function. This may cause that the built-in
function will go on the iterations even if the height has been greater than $M$.

In our implementations of PSLQ and IPSLQ, the same function uses the same technique for fairness.
According to Table \ref{tab:MPSLQvsPSLQ}, the IPSLQ  algorithm is obviously
faster than the PSLQ algorithm. Meanwhile the ratio between $T_{PSLQ}$ and $T_{IPSLQ}$ seems to
get larger and larger with increasing $d$.

\newcommand{\noopsort}[1]{}

\end{document}